\documentclass[11pt]{article}
\usepackage{jcappub}
\usepackage{graphicx} 
\usepackage{amsmath}

\makeatletter
\gdef\@fpheader{}
\makeatother

\newcommand{\sss}[1]{{\scriptscriptstyle{#1}}}

\newcommand{\negleft}{\negthinspace\left}
\newcommand{\dirac}[1]{\delta\negleft(#1\right)}
\newcommand{\diracb}[1]{\delta\negleft[#1\right]}
\newcommand{\step}[1]{\Theta\negleft( #1 \right)}
\newcommand{\stepb}[1]{\Theta\negleft[ #1 \right]}

\newcommand{\cte}{\mathrm{C}^{\mathrm{te}}}

\newcommand{\ud}{\mathrm{d}}

\newcommand{\uv}{\mathrm{v}}

\newcommand{\up}{\mathrm{p}}
\newcommand{\uf}{\mathrm{f}}

\newcommand{\uini}{\mathrm{ini}}
\newcommand{\ucur}{\mathrm{cur}}
\newcommand{\uprod}{\mathrm{prod}}

\newcommand{\ux}{\mathrm{x}}

\newcommand{\uT}{\mathrm{T}}

\newcommand{\calJ}{\mathcal{J}}
\newcommand{\calP}{\mathcal{P}}
\newcommand{\calF}{\mathcal{F}}
\newcommand{\calN}{\mathcal{N}}
\newcommand{\barcalN}{\bar{\mathcal{N}}}
\newcommand{\calV}{\mathcal{V}}

\newcommand{\calNini}{\calN_\uini}
\newcommand{\calVini}{\calV_\uini}
\newcommand{\barcalNini}{\barcalN_\uini}

\newcommand{\rhoinf}{\rho_\infty}

\newcommand{\Z}{Z}
\newcommand{\N}{N}
\newcommand{\Nstar}{\N_*}
\newcommand{\Tx}{T_\ux}
\newcommand{\Tcur}{T_\ucur}

\newcommand{\ellv}{\ell_\uv}
\newcommand{\ellp}{\ell_\up}
\newcommand{\ellini}{\ell_\uini}
\newcommand{\ellT}{\ell_{\sss{\uT}}}
\newcommand{\ellstar}{\ell_*}

\newcommand{\gammad}{\gamma_\ud}
\newcommand{\gammav}{\gamma_\uv}

\newcommand{\zini}{z_\uini}
\newcommand{\zv}{z_\uv}
\newcommand{\tini}{t_\uini}
\newcommand{\tcur}{t_\ucur}
\newcommand{\tstar}{t_*}
\newcommand{\aini}{a_\uini}
\newcommand{\Jacob}{\calJ}
\newcommand{\taup}{\tau_\up}
\newcommand{\tpv}{t_{\up\uv}}

\def\setR{\mathbb{R}}
\def\setC{\mathbb{C}}

\newcommand{\GN}{G_{_\mathrm{N}}}

\title{A Boltzmann treatment for the vorton excess problem}

\author[a]{Patrick Peter}
\author[b]{and Christophe Ringeval}

\affiliation[a]{${\cal G}\setR\varepsilon\setC{\cal O}$ -- Institut
d'Astrophysique de Paris, UMR7095 CNRS, Universit\'e Pierre \& Marie Curie,
98 bis boulevard Arago, 75014 Paris, France}

\affiliation[b]{Centre for Cosmology, Particle Physics and Phenomenology,
  Institute of Mathematics and Physics, Louvain University, 2 Chemin
  du Cyclotron, 1348 Louvain-la-Neuve (Belgium)}

\emailAdd{peter@iap.fr}
\emailAdd{christophe.ringeval@uclouvain.be}

\abstract{We derive and solve a Boltzmann equation governing the
  cosmological evolution of the number density of current carrying
  cosmic string loops, whose centrifugally supported equilibrium configurations
  are also referred to as {\sl vortons}. The phase space
  is three-dimensional and consists of the time variable, the loop
  size, and a conserved quantum number. Our approach includes
  gravitational wave emission, a possibly finite lifetime for the
  vortons and works with any initial loop distribution and for any
  loop production function. We then show how our results generalize
  previous approaches on the vorton excess problem by tracking down
  the time evolution of the various sub-populations of
  current-carrying loops in a string network.}


\begin{document}
\maketitle

\section{Introduction}

Cosmic strings are one of the expected, and actively searched,
signatures of early universe physics in
cosmology~\cite{Fraisse:2007nu, Takahashi:2008ui, Hindmarsh:2009qk,
  Hindmarsh:2009es, Regan:2009hv, Yamauchi:2010vy, Landriau:2010cb,
  Bevis:2010gj, Ringeval:2010ca, Tashiro:2012nb, Cai:2012zd,
  Ringeval:2012tk}. Originally, they were considered as line-like
topological defects produced during the phase transitions associated
with the spontaneous breakdown of Grand Unified
symmetries~\cite{Kirzhnits:1972, Kibble:1976}. More recently, it was
realized that very similar objects could also be generated within
Superstrings Theory and formed at the end of brane
inflation~\cite{Burgess:2001fx, Sarangi:2002yt, Dvali:2003zj,
  Jones:2003da, Davis:2005, Sakellariadou:2006qs,
  Sakellariadou:2008ie, Copeland:2009ga, Sakellariadou:2009ev}. In
their simplest version, cosmic strings exhibit Lorentz invariance
along the string worldsheet and the two prototypical examples are the
Abelian--Higgs string and the Nambu--Goto string. Both of these models
are simple enough to allow for numerical simulations of cosmic strings
evolution in Friedmann--Lema\^{\i}tre--Robertson--Walker (FLRW)
spacetimes~\cite{Albrecht:1989, Bennett:1989, Bennett:1990,
  Allen:1990, Vincent:1996rb, Vincent:1998, Moore:2002,
  Ringeval:2005kr, Hindmarsh:2008dw}. These simulations have shown
that, due to their so-called scaling regime, long strings never
dominate the energy density of the Universe and evolve as $\rhoinf
\propto 1/t^2$, instead of the naively expected $1/a^2$ behaviour for
one-dimensional objects ($t$ being the cosmic time and $a$ the scale
factor). In this regime, the unique model parameter is the string
energy density per unit length, $U$, which is constrained by Cosmic
Microwave Background (CMB) measurements. For the Abelian--Higgs string
network, Ref.~\cite{Urrestilla:2011gr} reports a two-sigma upper bound
$\GN U< 4.2 \times 10^{-7}$, $\GN$ being the Newton constant. For
Nambu--Goto strings, the scaling evolution takes place via the
emission of loops whereas the Abelian--Higgs string scaling seems to
mainly rely on particle emission~\cite{Moore:2002,
  Hindmarsh:2008dw}. In addition to be a priori model-dependent, the
expected cosmological distribution of loops is a matter of debate. As
shown in Refs.~\cite{Vanchurin:2005yb, Ringeval:2005kr}, a copious
number of small loops seen in Nambu--Goto simulations can be linked to
some relaxation effects coming from the initial conditions, and as
such they are not expected to persist over cosmological
timescales. However, in addition to these transient loops,
Ref.~\cite{Ringeval:2005kr} was the first to exhibit a sub-population
of loops in scaling, i.e having an energy density evolution similar to
the one of the long strings. Their distribution was also found to
match the analytic and independent findings of
Refs.~\cite{Polchinski:2006ee, Dubath:2007mf, Rocha:2007ni}. Contrary
to the former, these loops are expected to persist during the
cosmological expansion. Those results were recovered in
Refs.~\cite{Vanchurin:2005pa, Martins:2005es, Olum:2006ix}, albeit
with some differences in the distribution, while the authors of a more
recent simulation presented in Ref.~\cite{BlancoPillado:2011dq}
arguably claim that all loops but the Hubble-sized ones (the so-called
Kibble loops) should ultimately disappear\footnote{The string
  simulation of Ref.~\cite{BlancoPillado:2011dq} has currently the
  longest dynamical time, but at the expense of having the poorest
  resolution. Their discretized strings having only two points per
  initial correlation length as opposed to twenty for
  Ref.~\cite{Ringeval:2005kr} and hundreds for
  Ref.~\cite{Martins:2005es}. Among others, a low resolution has
  precisely the effect of artificially cutting the loop distribution
  at small scales (see Fig.~2 in Ref.~\cite{Ringeval:2005kr}).} (see
Ref.~\cite{Ringeval:2010ca} and references therein for a review).

Even if we could infer the exact loop production function from the
string simulations, free of any numerical uncertainties, one would
have to keep in mind that the dynamics of the smallest length scales
still involves physical processes that are not included in numerical
simulations. For instance, gravitational wave emission and
gravitational backreaction are crucial ingredients which require the
use of analytic and semi-analytic models to be dealt
with~\cite{Austin:1993rg, Martins:2000cs, Copeland:2009dk,
  Martins:2009hj, Lorenz:2010sm, Pourtsidou:2010gu,
  Avgoustidis:2011ax, Vanchurin:2011hm}. In particular, based on the
loop production function of Refs.~\cite{Polchinski:2006ee,
  Dubath:2007mf}, the Boltzmann approach of Ref.~\cite{Rocha:2007ni}
has been extended in Ref.~\cite{Lorenz:2010sm} to include the effects
of the initial conditions and gravitational backreaction into the
evolution of Nambu--Goto cosmic string loops. A Boltzmann treatment
for loops consists in describing their evolution in phase space,
e.g. $(t,\ell)$ if $\ell$ denotes the loop size, in order to extract
their number density distribution $n(t,\ell)$. Such a method has first
been used in Ref.~\cite{Copeland:1998na} to study the generation of
long strings from networks having only loops but also for the search of
cosmic string signatures~\cite{Leblond:2009fq, Tashiro:2012nv}.

In this paper, we go one step further and present a new Boltzmann
treatment dealing with the so-called current carrying cosmic string
loops, a class of uttermost importance for
cosmology~\cite{Witten:1984eb}. These objects naturally appear as soon
as the strings exhibit coupling with bosonic or fermionic
particles~\cite{Carter:1989dp,Peter:1992dw, Carter:1999hx,
  Ringeval:2000kz, Ringeval:2001xd, Carter:2003fb}. This results into
the breaking of Lorentz invariance along the worldsheet thereby allowing
the existence of centrifugally supported loops called
vortons~\cite{Davis:1988ip, Davis:1988ij}. If these objects are stable
over cosmological timescales~\cite{Carter:1989xk, Carter:1993wu,
  Davis:1999ec, Peter:2000sw} they could come to dominate the energy
budget of the Universe and this has been used in
Refs.~\cite{Brandenberger:1996zp, Carter:1999an} to severely
constrain the energy scale at which they may be formed. In these
papers, it was nevertheless assumed that most of the loops, which can
potentially become vortons at later times, are formed once and for all
at the string forming time. In view of the new numerical results
described above, one may wonder whether more drastic constraints are
not in place when one takes into account the loops produced during the
cosmological evolution of the string network.

In the following, we tackle this problem and solve a Boltzmann
equation describing the production and decay of current carrying
loops. Our approach includes gravitational radiation, finite life-time
vortons with any initial distributions and any loop production
functions. After having presented our working hypothesis and written
down the Boltzmann equation, we first present the Liouville solution
in Sec.~\ref{sec:liouevol}, i.e. without a loop production
function. This allows us to physically discuss the different
population of loops present at any time. In Sec.~\ref{sec:boltzevol},
we solve the complete Boltzmann equation, in presence of a loop
production function. This is our main result and it is
presented in Eqs.~(\ref{eq:liousol-}), (\ref{eq:boltzsol+}) and
(\ref{eq:boltzsol0}). In Sec.~\ref{sec:vabund}, we generalize the
results of Ref.~\cite{Brandenberger:1996zp} to the case of perfectly
stable vortons created from any initial loop distribution. This allows
us to track down the time evolution of the so-called proto-vortons,
doomed loops and vortons. We conclude in the last section.

\section{Boltzmann approach}
\label{sec:evol}

\subsection{Assumptions}

Our main assumption is that the loop production mechanism remains
similar to the Nambu--Goto case even though the strings are now
current-carrying. This can be motivated from the macroscopic covariant
formalism of Carter~\cite{Carter:1989dp, Carter:2000wv} as the
existence of a current can be viewed as modifying the Nambu--Goto
relation $U=T$ (where $T$ is the string tension) into a more general
equation of state $U(T)$. Provided the currents are not too strong, one
has $U\simeq T$~\cite{Peter:1992dw, Carter:1994hn} and the string
dynamics remain similar to the Nambu--Goto one. Let us notice that
some cosmic superstring models clearly violate this assumption, as for
instance those developing $Y$-junctions~\cite{Copeland:2006eh,
  Copeland:2006if, Copeland:2007nv, Pourtsidou:2010gu}.

Current-carrying loops are assumed to be created with two conserved
quantum and topological
numbers $\N$ and $\Z$ associated with the existence of a
conserved current flowing along the
string~\cite{Carter:1990sm}. Following
Ref.~\cite{Brandenberger:1996zp}, we assume $\Z \simeq \N$ with
\begin{equation}
\label{eq:Nformed}
\left. \N \right|_{\uprod} \simeq \sqrt{\dfrac{\ell}{\lambda}}\,.
\end{equation}
Here $\ell$ is the physical length of a loop at formation and
$\lambda$ the typical wavelength of the carrier field fluctuations,
which is considered to be constant (originating from microphysics,
it can be assumed to be of the order of the Compton wavelength
of the condensed particle). This expression should be valid
provided $\lambda$ remains much smaller than the mean interstring
distance. In this paper, we will make no assumptions on the explicit
shape of the loop production function and simply call it
$\calP(\ell,t)$.  As a well-motivated simple example, one can pick
that of Refs.~\cite{Polchinski:2006ee, Dubath:2007mf, Rocha:2007ni} for
small loops, adding a Dirac distribution centered at the mean
inter-string distance for the Kibble loops.

\subsection{Evolution in phase space}

\begin{figure}
\begin{center}
\includegraphics[width=0.8\textwidth]{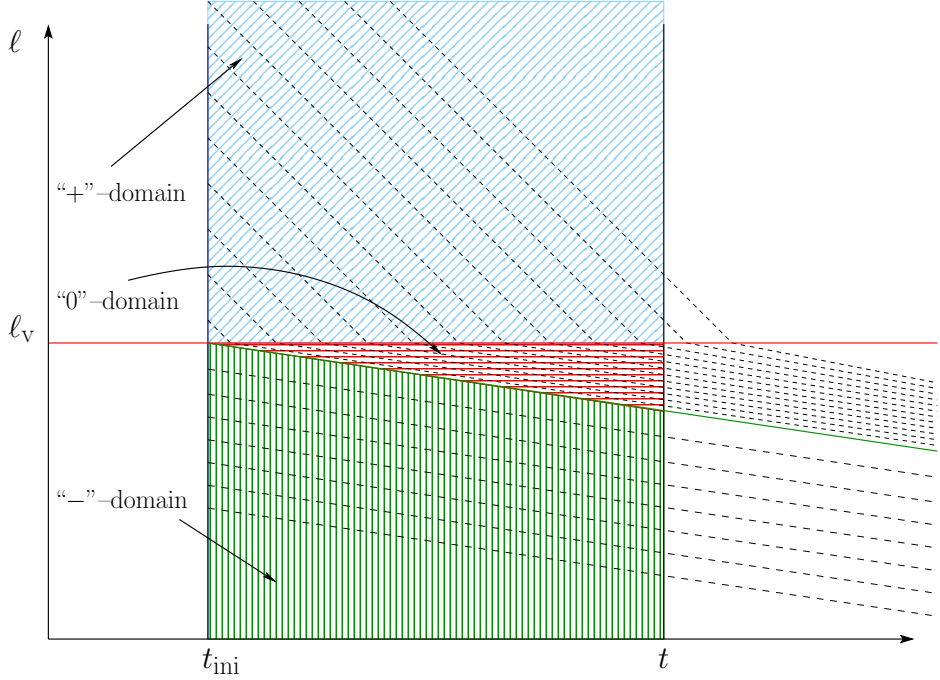}
\caption{Characteristics $\xi(\ellini,t)$ of the Liouville operator at fixed
  $\N$ in the plane $(t,\ell)$. At any time $t$, the Cauchy problem is
  solved by specifying initial conditions on the surface $t=\tini$ and
  propagating them along the characteristics in the ``+'' and
  ``--'' regions. Notice that the solution in the ``0'' region is
  determined by specifying initial condition on the line $\ell=\ellv$,
  which can still be deduced from $t=\tini$ after evolution through
  the ``+'' domain.}
\label{fig:domains}
\end{center}
\end{figure}

Denoting by $n(\ell,\N, t)$ the number density distribution of cosmic
string loops of size $\ell$,  and conserved quantum number $\N$ at
cosmic time $t$, we can write
\begin{equation}
\label{eq:dpart}
\dfrac{\ud}{\ud t} \left(a^3 \Jacob \dfrac{\partial^2 n}{\partial \ell
  \partial \N} \right) = a^3 \Jacob \calP(\ell,t) \dirac{\N -
  \sqrt{\dfrac{\ell}{\lambda}}} ,
\end{equation}
where $\calP(\ell,t)$ is the loop production function discussed above,
which is a collision term from the Boltzmann point of view, $a^3$ accounts
for space-time expansion while $\Jacob(\ell,t)$ encodes any additional
phase space distortion. The Dirac distribution ensures that all loops
freshly formed carry the expected quantum numbers provided by
Eq.~(\ref{eq:Nformed}). Due to gravitational radiation, a
current-carrying loop of size $\ell(t)$ shrinks at a constant rate
until it eventually becomes a vorton, i.e. a state centrifugally supported by
its current. Its fate depends on the microscopic model under
scrutiny. For the sake of generality, we are assuming that vortons can
decay by an unspecified mechanism at a constant rate $\gammav$, the
completely stable situation corresponding to $\gammav \rightarrow
0$. As a result, we model the evolution of $\ell(t)$ for a loop of
conserved number $\N$ by
\begin{equation}
\label{eq:ellevol}
\dfrac{\ud \ell}{\ud t} = -\gammad \step{\ell - \ellv} - \gammav
\step{\ellv-\ell},
\end{equation}
where $\step{x}$ denotes the Heaviside function. The gravitational
wave emission rate, $\gammad$, is assumed to be the same as for
Nambu-Goto strings
\begin{equation}
\label{eq:gammad}
  \gammad \simeq \Gamma \GN U,
\end{equation}
where $\Gamma$ is a constant of order $10^2$ which depends on the string
structure, geometry and dynamics~\cite{Vilenkin:1981, PhysRevD.45.1898}.
Having in mind the limit $\gammav/\gammad\to 0$ for the actual vorton situation,
one can safely assume $\gammav/\gammad\ll 1$ in numerical evaluations.
The quantity
$\ellv(\N)$ is the physical length of vortons typically given
by~\cite{Brandenberger:1996zp, Carter:1999an}
\begin{equation}
\label{eq:ellv}
\ellv(\N) =\sqrt{\dfrac{\N \Z}{U}} \simeq \dfrac{\N}{\sqrt{U}}\,.
\end{equation}
In the phase space $(t,\ell,\N)$, Eq.~(\ref{eq:ellevol}) does not
preserve volume. All loops contained in a given interval $\Delta
\ellini$ evolve according to Eq.~(\ref{eq:ellevol}) into a smaller $\Delta
\ell(t)$ interval once $\ell(t)<\ellv$. We are therefore in presence
of a generalized Liouville evolution and $\Jacob(\ell,t)$ is the
Jacobian of the transformation mapping $\ellini$ to
$\ell(t)=\xi(\ellini,t)$, i.e.
\begin{equation}
\Jacob(\ell,t) = \left| \dfrac{\partial \xi(\ellini,t)}{\partial \ellini}\right|.
\end{equation}
The function $\xi(\ellini,t)$ is the characteristic curve starting at
$\ellini$ and solution of Eq.~(\ref{eq:ellevol}) (see
Fig.~\ref{fig:domains}). Concerning the conserved number $\N$, by
definition we have $\ud \N(t)/\ud t = 0$ along any loop trajectory and
there is no phase space volume distortion along the $\N$
direction. Combining Eqs.~(\ref{eq:dpart}) and (\ref{eq:ellevol}) gives
the Boltzmann equation
\begin{equation}
\label{eq:boltzell}
\begin{aligned}
  \dfrac{\partial}{\partial t} \left[a^3 \Jacob(\ell,t) \dfrac{\partial^2
      n}{\partial \ell \partial \N} \right] & - \left[\gammad \step{\ell -
      \dfrac{\N}{\sqrt{U}}} + \gammav
    \step{\dfrac{\N}{\sqrt{U}}-\ell}\right] \dfrac{\partial}{\partial
    \ell}\left[a^3 \Jacob(\ell,t) \dfrac{\partial^2 n}{\partial \ell \partial \N}
    \right] \\ & = a^3 \Jacob(\ell,t) \calP(\ell,t) \dirac{\N -
    \sqrt{\dfrac{\ell}{\lambda}}}.
\end{aligned}
\end{equation}

In order to make contact with previous works for Nambu-Goto
strings~\cite{Lorenz:2010sm}, one can also express the loop sizes $\ell$ in
units of the cosmic time $t$. For this purpose, we will also use in
the following the variable $\gamma$ and the density function $\calF$
defined as
\begin{equation}
\label{eq:calFdef}
  \gamma(\ell,t) \equiv \dfrac{\ell}{t}\,, \qquad \calF \equiv
  \Jacob \dfrac{\partial^2 n}{\partial \ell \partial \N}\,.
\end{equation}
For a given value of $\N$, Eq.~(\ref{eq:boltzell}) exhibits two
typical length scales, one is $\ellv(\N)$ already given by
Eq.~(\ref{eq:ellv}) and the other is
\begin{equation}
\label{eq:ellpdef}
\ellp(\N) \equiv \lambda N^2\,,
\end{equation}
which gives the size of a loop with conserved number $\N$ at
production time. For $\N=0$, we have $\ellv=\ellp=0$, otherwise
\begin{equation}
\ellv(N) \ll \ellp(N)\,,
\end{equation}
assuming $\sqrt{U} \gg \lambda^{-1}$. In the following, we first discuss
the Liouville equation, i.e. Eq.~(\ref{eq:boltzell}) without the
production term, before solving the full Boltzmann equation.

\subsection{Liouville evolution}

\label{sec:liouevol}

In order to get intuition on the solution, we solve first the
Liouville equation, i.e. without including the loop production. Under
this assumption, one should recover the results of
Ref.~\cite{Brandenberger:1996zp} since all the produced vortons come
only from the initial loop distribution (see also Sec.~\ref{sec:vabund}).

\subsubsection{Jacobian}

In order to determine $\Jacob(\ell,t)$, the function $\xi(\ellini,t)$
is needed. Integrating Eq.~(\ref{eq:ellevol}) gives
\begin{equation}
\label{eq:xi}
\begin{aligned}
\xi(\ellini,t) & = \step{\ellv-\ellini}\left(\ellini - \gammav t
\right) + \step{\ellini-\ellv}\bigg\{ \left(\ellini - \gammad t
\right) \step{\ellini - \ellv -\gammad t} \\ & + \left[ \ellv -
  \gammav t + \dfrac{\gammav}{\gammad} \left(\ellini - \ellv \right)
  \right] \step{\gammad t +\ellv - \ellini}  \bigg\},
\end{aligned}
\end{equation}
from which we deduce that
\begin{equation}
\begin{aligned}
\dfrac{\partial \xi}{\partial \ellini}(\ellini,t) = \step{\ellv
  -\ellini} + \step{\ellini -\ellv} \left[
  \step{\ellini-\ellv -\gammad t} + \dfrac{\gammav}{\gammad}
  \step{\gammad t +\ellv - \ellini} \right],
\end{aligned}
\end{equation}
where we have set $\tini=0$.

The Jacobian $\Jacob(\ell,t)$ is obtained by taking the modulus of the
above equation, up to a change of variables $(\ellini,t) \rightarrow
(\ell,t)$. This requires the inversion of Eq.~(\ref{eq:xi}) to get
$\ellini = \xi^{-1}(\ell,t)$. After some algebra, the final expression
simplifies considerably and reads
\begin{equation}
\label{eq:jacobstep}
\Jacob(\ell,t) = \step{\ell-\ellv} + \step{\ellv -\ell} \left[
  \step{\ellv - \ell - \gammav t} + \dfrac{\gammav}{\gammad} \step{\ell
    + \gammav t - \ellv} \right],
\end{equation}
which merely reflects the changes of slopes of the decaying processes
(see Fig.~\ref{fig:domains} discussed below).
At any time $t$ and for any value of $\N$, the Jacobian is unity
for all loop sizes satisfying $\ell > \ellv(N)$. It then switches to
the value $\gammav/\gammad$ in the domain $\ellv
- \gammav t <\ell < \ellv$ and returns to one again for $\ell < \ellv -
\gammav t$. Fig.~\ref{fig:domains} shows these three
domains in the $(t,\ell)$ plane, respectively denoted as ``+'', ``0''
and ``--''. Their origin lies in the various histories a $\ellini$-sized loop
starting at $t=\tini$ may have. From Eq.~(\ref{eq:xi}), all loops
starting with $\ellini < \ellv$ (``--'' domain) decay with a rate
given by $\gammav$  as being already centrifugally supported. The loop
number density distribution per $\Delta \ell$ interval is thus
preserved and the Jacobian is unity. At any time $t$, all loops having
$\ell>\ellv$ are necessarily coming from $\ellini > \ellv$. Since they
never crossed $\ellv$, they have shrunk only by gravitational
radiation, again at a constant rate given by $\gammad$, the Jacobian
is also equal to unity. Only the loops in the ``0'' domain have a two
steps history. They started with $\ellini > \ellv$, then radiated
gravitational waves until they became vortons by hitting
$\ell=\ellv$. Once vortonized, their energy emission rate being
$\gammav \ll \gammad$, they start accumulating in this region and the
Jacobian takes the value $\gammav/\gammad \ll 1$.

\subsubsection{Loop number density distribution}

We are now in the position to solve Eq.~(\ref{eq:boltzell}) without
the collision term, i.e. in absence of loop production. This is a
first order differential equation whose characteristics are given by
Eq.~(\ref{eq:xi}).

\begin{figure}
\begin{center}
\includegraphics[width=0.7\textwidth]{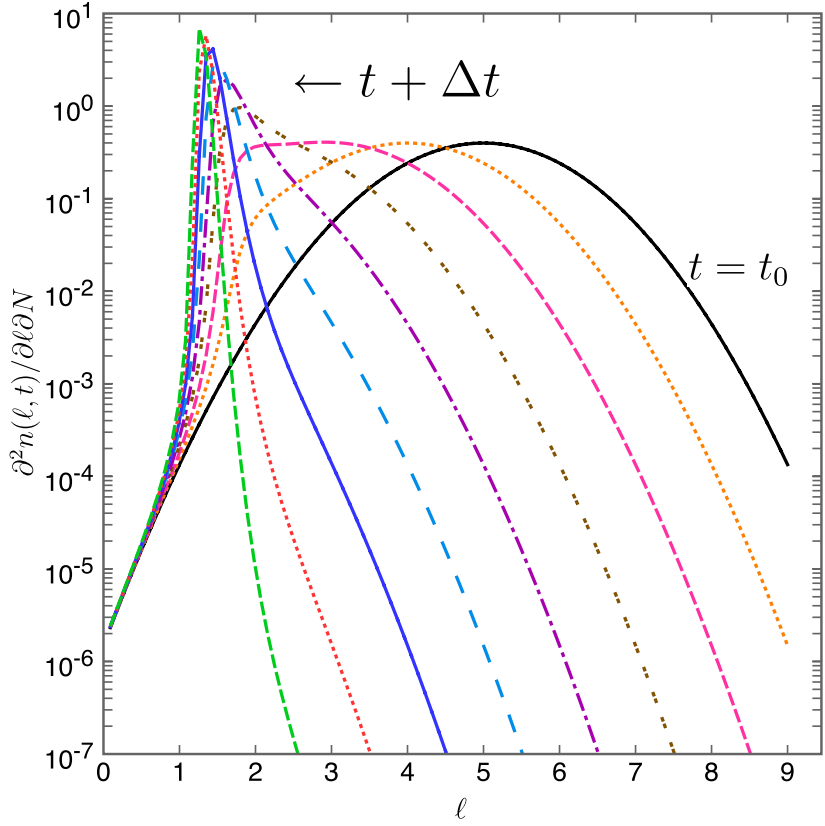}
\caption{Liouville time evolution of the loop distribution given by
  Eqs.~(\ref{eq:liousol+}), (\ref{eq:liousol0}) and
  (\ref{eq:liousol-}) with $\gammad=1$. At $t=\tini$, we assume a
  gaussian initial distribution centered at $\ell=\ell_\up=5$ with a
  standard deviation of $\sigma_\up=1$ (full black line). We also
  assumed perfectly stable vortons, so that $\gammav=0$, and we
  smoothed the step function appearing in Eq.(\ref{eq:ellevol}) with
  $\Theta (\ell -\ellv)\to \left\{ 1+\tanh \left[(\ell-\ellv)/\lambda
    \right] \right\}/2$, and $\lambda=0.5$: this is the reason why the
  distribution is non-vanishing for values of $\ell$ smaller than
  $\ellv$. As time evolves, the distribution is shifted to the left,
  where it starts rapidly feeling the vorton characteristic scale
  $\ellv$, here numerically set to $2$. Each curve represents the
  distribution one unit of time after the previous curve, i.e., in the
  units used, the initial gaussian is for $t=0$, and then $t=1, 2,
  \dots,8$ (orange dotted, pink dashed, and so on, respectively).  As
  time passes, the distribution evolves to a very sharp peak located
  around $\ellv$, which gives a Dirac distribution $\delta
  (\ell-\ellv)$ in the limit $\lambda\to 0$.}
\label{fig:distrib}
\end{center}
\end{figure}

According to the values of $\ell$, it is convenient to introduce the
new coordinates $(\tau,x)$ and $(\tau,y)$ defined as
\begin{equation}
\label{eq:newcoords}
\begin{aligned}
\forall \ell > \ellv(N) & \quad \Rightarrow \quad \tau \equiv t, \quad x
\equiv \gammad t + \ell \, , \\ \forall \ell < \ellv(N) & \quad
\Rightarrow \quad \tau \equiv t, \quad y \equiv \gammav t + \ell \,.
\end{aligned}
\end{equation}
In terms of these variables, the Liouville equation,
i.e. Eq.~(\ref{eq:boltzell}) with $\calP(\ell,t)=0$, takes the simple form
\begin{equation}
\label{eq:conservation}
\dfrac{\partial (a^3 \calF)}{\partial \tau} = 0\,.
\end{equation}
The solutions are
\begin{equation}
\label{eq:liousol}
\begin{aligned}
a^3(t) \calF(t,\ell>\ellv,N) & = I(\ell + \gammad t,N),\\
a^3(t) \calF(t,\ell<\ellv,N) & = J(\ell + \gammav t,N),
\end{aligned}
\end{equation}
where $I(\cdot,\cdot)$ and $J(\cdot,\cdot)$ are two arbitrary
functions still to be determined. Given the loop distribution at some
initial time $\tini$,
\begin{equation}
\label{eq:calNini}
\dfrac{\partial^2 n}{\partial \ell \partial \N} (\tini,\ell,\N) = \calNini(\ell,\N)\,,
\end{equation}
one can determine these two functions in their respective domain of
definition. Plugging Eq.~(\ref{eq:liousol}) into Eq.~(\ref{eq:calNini})
and making use of Eqs.~(\ref{eq:calFdef}) and (\ref{eq:jacobstep}),
all evaluated at $t=\tini$ gives
\begin{equation}
I(z>\zini,\N) = \aini^3 \calNini(z-\gammad \tini,\N),
\end{equation}
where $\zini \equiv \ellv + \gammad \tini$. This expression completely
fixes the solution inside the ``+'' domain as
\begin{equation}
\label{eq:liousol+}
\begin{aligned}
a^3(t) \dfrac{\partial^2 n}{\partial \ell \partial \N} (t,\ell,\N)  =
\aini^3 \calNini\negleft[\ell + \gammad(t-\tini),\N\right], & \\
 \ell > \ellv(\N). &
\end{aligned}
\end{equation}
According to the discussion of the previous section, the solution in
the ``0'' domain is uniquely determined by propagating this solution
to $\ell=\ellv$, at all times. The matching condition
implies\footnote{The conserved quantity is $a^3 \calF$ which already
  includes the Jacobian factor, see Eq.~(\ref{eq:conservation}).}
\begin{equation}
\aini^3 \calNini\negleft[\ellv + \gammad(t - \tini),\N \right] = J(\ellv
+ \gammav t,\N),
\end{equation}
such that
\begin{equation}
J(z>\zv,\N) = \aini^3 \calNini\negleft[\ellv +
  \dfrac{\gammad}{\gammav}\left(z -\ellv\right) - \gammad \tini,\N \right].
\end{equation}
Here, we have defined $\zv \equiv \ellv + \gammav \tini$. Plugging
this expression back into Eq.~(\ref{eq:liousol}), restricted on $\ell
+ \gammav t > \zv$ (and $\ell<\ellv$) completely fixes the solution inside
the ``0'' domain
\begin{equation}
\label{eq:liousol0}
\begin{aligned}
a^3(t) \dfrac{\partial^2 n}{\partial \ell \partial \N}(t,\ell,\N) = \aini^3
\dfrac{\gammad}{\gammav} \calNini\negleft[\ellv +
  \dfrac{\gammad}{\gammav}\left(\ell - \ellv \right) + \gammad \left(t
  - \tini\right),\N \right],& \\
\ellv(\N)- \gammav(t-\tini) < \ell< \ellv(\N) \, . &
\end{aligned}
\end{equation}
At that point, $J(z,\N)$ is not yet determined for $z<\zv$. However,
this part is exactly the ``--'' domain for which the initial
conditions are fixed at $t=\tini$ and
\begin{equation}
\label{eq:J-}
J(z<\zv,\N) = \aini^2 \calNini(z - \gammav \tini,\N).
\end{equation}
The solution finally reads
\begin{equation}
\label{eq:liousol-}
\begin{aligned}
a^3(t) \dfrac{\partial^2 n}{\partial \ell \partial \N} (t,\ell,\N)  =
\aini^3 \calNini\negleft[\ell + \gammav(t-\tini),\N\right], & \\
 \ell < \ellv(\N) - \gammav(t - \tini). &
\end{aligned}
\end{equation}
The loop distribution at all times $t>\tini$, length $\ell$ and
conserved quantum number $\N$ is then given by the set of
Eqs.~(\ref{eq:liousol+}), (\ref{eq:liousol0}) and
(\ref{eq:liousol-}). Although the solutions in region ``+'' and ``--''
could have been intuitively guessed as they only shift the initial
loop distribution according to the gravitational and vortonic energy
emission rates, Eq.~(\ref{eq:liousol0}) shows that loops tend to
accumulate into the region ``0''.

This is exemplified on Fig.\ref{fig:distrib}, showing the time
evolution of an initially gaussian distribution and producing
perfectly stable vortons, case to which we now turn.

\subsubsection{Perfectly stable vortons}

In the limit of perfectly stable vortons, i.e. $\gammav \rightarrow
0$, Eq.~(\ref{eq:liousol0}) is singular and the loop density
distribution becomes infinite but into a phase space region which
becomes infinitely small (see Fig.~\ref{fig:domains}). This is not
problematic as the physical quantity of interest is the total number
of loop in that region. We can therefore define the number of
``vortons'' having a charge $\N$, at time $t$ by
\begin{equation}
\label{eq:calVdef}
\calVini(t,\N) \equiv \int_{\ellv - \gammav(t-\tini)}^{\ellv}
\negthickspace \dfrac{\partial^2 n}{\partial \ell \partial
  \N}(t,\ell,\N) \, \ud \ell \,,
\end{equation}
where ``ini'' is a reminder that, in absence of loop production, they
can only come from the initial loop distribution. Using
Eq.~(\ref{eq:liousol0}), defining for each time slice the new variable
\begin{equation}
\label{eq:sigmadef}
\sigma \equiv \dfrac{\gammad}{\gammav} \left(\ell - \ellv \right)  + \gammad(t
  - \tini),
\end{equation}
the integral reduces to
\begin{equation}
\label{eq:vortdensini}
\calVini(t,\N) = \dfrac{\aini^3}{a^3(t)} \int_{0}^{\gammad(t
  - \tini)} \calNini(\ellv + \sigma,\N) \ud \sigma\,,
\end{equation}
which is well-defined and finite. For an homogeneous initial
distribution ($\calNini=\cte$) one recovers the intuitive result that
the number of vortons grows linearly with time as $a^3 \calVini
\propto \gammad (t-\tini)$.

We conclude that, in the limit of perfectly stable vortons $\gammav
\rightarrow 0$, the loop density distribution reads, for all $\ell$,
\begin{equation}
\label{eq:lioudens}
\begin{aligned}
\lim_{\gammav \rightarrow 0} a^3(t) \dfrac{\partial^2 n}{\partial \ell \partial
  \N}(t,\ell,\N) & = \step{\ellv- \ell} \aini^3  \calNini(\ell,\N) \\ &
+ \step{\ell-\ellv} \aini^3  \calNini\negleft[\ell +
  \gammad \left( t-\tini \right),\N \right] \\
&+ \dirac{\ell - \ellv} \aini^3  \int_{0}^{\gammad(t
  - \tini)} \calNini(\ellv + \sigma,\N) \ud \sigma\,.
\end{aligned}
\end{equation}

\subsection{Boltzmann evolution with loop production}

\label{sec:boltzevol}

We now solve Eq.~(\ref{eq:boltzell}) in full generality by separating
the phase space into the same three domains.

\subsubsection{Formal solution}

For all $\ell>\ellv(N)$, Eq.~(\ref{eq:boltzell}) can be recast in
terms of the variable $(\tau,x)$ introduced in
Eq.~(\ref{eq:newcoords}) as
\begin{equation}
\dfrac{\partial\left(a^3 \calF\right)}{\partial \tau} =
a^3(\tau) \calP(x-\gammad \tau,\tau) \dirac{\N-\sqrt{\dfrac{x -
      \gammad \tau}{\lambda}}},
\end{equation}
the Jacobian being unity there. The delta function has an argument
depending on $\tau$ such that it simplifies to
\begin{equation}
\dirac{\N-\sqrt{\dfrac{x - \gammad \tau}{\lambda}}} = \dfrac{2
  \ellp(\N)}{\N \gammad} \, \diracb{\tau - \taup(x,\N)},
\end{equation}
where $\ellp(\N)$ is given by Eq.~(\ref{eq:ellpdef}) and
\begin{equation}
\taup(x,\N) \equiv \dfrac{x - \ellp(\N)}{\gammad}\,.
\end{equation}
The loop distribution in the ``+'' domain is then solution of
\begin{equation}
\dfrac{\partial \left(a^3 \calF \right)}{\partial \tau} = 2
a^3[\taup(x,\N)] \, \dfrac{\ellp(\N)}{\gammad \N} \,
\calP[\ellp(\N),\taup(x,\N)] \, \diracb{\tau - \taup(x,\N)},
\end{equation}
and reads
\begin{equation}
a^3 \calF(\tau,x,\N) = I(x,\N) + 2 a^3[\taup(x,\N)] \,
\dfrac{\ellp(\N)}{\gammad \N} \, \calP[\ellp(\N),\taup(x,\N)] \,
\stepb{\tau - \taup(x,\N)}.
\end{equation}
As for the Liouville equation, $I(x,\N)$ is a unknown function that
has to be determined from the initial conditions. In terms of the
original variables, the previous equation reads
\begin{equation}
\label{eq:largesol}
\begin{aligned}
a^3 \dfrac{\partial^2 n}{\partial \ell \partial \N}(t,\ell,\N) & =
I(\ell + \gammad t,\N) \\ & + \dfrac{2 \ellp}{\gammad \N} \,
a^3\negleft(t + \dfrac{\ell - \ellp}{\gammad} \right) \calP
\negleft(\ellp,t+\dfrac{\ell-\ellp}{\gammad} \right)
\step{\ellp - \ell}, \\ & & \ell > \ellv(\N).
\end{aligned}
\end{equation}

For all $\ell<\ellv(\N)$, we can similarly use the new variables
$(\tau,y)$ defined in Eq.~(\ref{eq:newcoords}) to solve
Eq.~(\ref{eq:boltzell}). In fact, since $\ellp \gg \ellv$ (or null),
the source term always vanish in that domain and the solution is the
same as in Sec.~\ref{sec:liouevol}, i.e.
\begin{equation}
\label{eq:smallsol}
\begin{aligned}
a^3 \Jacob(\ell,t) \dfrac{\partial^2 n}{\partial \ell \partial \N}(t,\ell,\N) =
J(\ell + \gammav t,\N),&\\ \ell < \ellv(\N),
\end{aligned}
\end{equation}
where the function $J(z,\N)$ has again to be determined from the
initial conditions. Notice that the Jacobian $\Jacob(\ell,t)$ is
either unity or equal to $\gammav/\gammad$ according to the domain
``--'' or ``0'', respectively. The formal solution is therefore given
by both Eqs.~(\ref{eq:largesol}) and (\ref{eq:smallsol}).

\subsubsection{Full solution from specified initial conditions}

As for the Liouville equation, we assume the loop distribution to be
known at some initial time $\tini$ and given by Eq.~(\ref{eq:calNini}).

For all $\ell > \ellv$, i.e. in the ``+'' domain,
Eq.~(\ref{eq:largesol}) evaluated at $t=\tini$ fixes the function
\begin{equation}
I(z>\zini,\N)  = \aini^3 \calNini(z-\gammad \tini,\N) - \dfrac{2
  \ellp}{\gammad \N} a^3 \negleft(\dfrac{z - \ellp}{\gammad} \right)
 \times \step{\tini + \dfrac{\ellp - z}{\gammad}}
\calP\negleft(\ellp,\dfrac{z-\ellp}{\gammad} \right),
\end{equation}
where $\zini = \ellv + \gammad \tini$. Notice that the Heaviside
function ensures that the argument of the scale factor $a(\cdot)$ is
always larger or equal than $\tini$. Plugging $I(z,\N)$, back into
Eq.~(\ref{eq:largesol}) gives the wanted result
\begin{equation}
\label{eq:boltzsol+}
\begin{aligned}
\begin{aligned}
a^3(t) \dfrac{\partial^2 n}{\partial \ell \partial \N}(t,\ell,\N) & = \aini^3
\calNini\negleft[\ell+\gammad(t-\tini),\N\right] + \dfrac{2
  \ellp}{\gammad \N} a^3\negleft(t - \dfrac{\ellp - \ell}{\gammad}
\right) \\ & \times
\calP\negleft(\ellp,t - \dfrac{\ellp-\ell}{\gammad}\right)
 \Big\{ \step{\ellp - \ell} - \stepb{\ellp - \ell - \gammad(t
  - \tini)} \Big\},
\end{aligned} & \\
\ell > \ellv(\N). &
\end{aligned}
\end{equation}

For all $\ell < \ellv(\N)$, we start from the solution
(\ref{eq:smallsol}). However, as in Sec.~\ref{sec:liouevol}, there are
two sub-cases corresponding to the domain ``0'' and ``--'' of
Fig.~\ref{fig:domains}.

On one hand, for $\ellv - \gammav(t-\tini) <\ell < \ellv$, we
determine the $J$ function by matching Eq.~(\ref{eq:smallsol}) to
Eq.~(\ref{eq:boltzsol+}) at $\ell = \ellv$ and for all times. One gets
\begin{equation}
\begin{aligned}
J(z & > \zv,\N)  = \aini^3 \calNini\negleft[\ellv +
  \dfrac{\gammad}{\gammav}(z - \ellv) - \gammad \tini,\N\right] +
\dfrac{2 \ellp}{\gammad \N} a^3\negleft(\dfrac{z - \ellv}{\gammav} -
\dfrac{\ellp - \ellv}{\gammad}\right) \\ \times & \calP
\negleft(\ellp,\dfrac{z-\ellv}{\gammav} - \dfrac{\ellp -
  \ellv}{\gammad} \right) \left[ \step{\dfrac{\ellp - \ellv}{\gammad}}
    - \step{\tini + \dfrac{\ellp - \ellv}{\gammad} -\dfrac{z
        -\ellv}{\gammav}}\right],
\end{aligned}
\end{equation}
where, as before, $\zv = \ellv + \gammav \tini$. This completely fixes
the solution in the ``0'' domain which reads
\begin{equation}
\label{eq:boltzsol0}
\begin{aligned}
\begin{aligned}
a^3(t) \dfrac{\partial^2 n}{\partial\ell \partial \N}(t,\ell,\N) & =
\aini^3 \dfrac{\gammad}{\gammav} \calNini\negleft[\ellv +
  \dfrac{\gammad}{\gammav} (\ell - \ellv) + \gammad(t - \tini),\N
  \right] \\& + \dfrac{2 \ellp}{\gammav \N} a^3\negleft(t -
\dfrac{\ellv - \ell}{\gammav} - \dfrac{\ellp - \ellv}{\gammad} \right)
\calP\negleft(\ellp,t - \dfrac{\ellv - \ell}{\gammav} - \dfrac{\ellp -
  \ellv}{\gammad}\right) \\
&\times \left\{\step{\ellp - \ellv} - \stepb{\ellp - \ellv +
  \dfrac{\gammad}{\gammav}(\ellv - \ell) - \gammad(t - \tini)} \right\},
\end{aligned} & \\
\ellv(\N) - \gammav(t-\tini) <\ell < \ellv(\N). &
\end{aligned}
\end{equation}
Notice that the Heaviside function $\step{ \ellp - \ellv} =1$ under
our hypothesis $\ellp \gg \ellv$.

On the other hand, when $\ell < \ellv - \gammav ( t-\tini)$, the
function $J(z<\zv,\N)$ is set by the initial loop distribution at $t =
\tini$. Since the source terms vanish in that region, combining
Eqs.~(\ref{eq:calNini}) and (\ref{eq:smallsol}) yields exactly
Eq.~(\ref{eq:J-}) for $J$. As a result, the loop distribution in that
domain is the same as the one derived from the Liouville operator in
Eq.~(\ref{eq:liousol-}).

To summarize, starting from the initial distribution of
Eq.~(\ref{eq:calNini}), in presence of a loop production
function, the loop number density distribution at time $t$, for all
$\ell$ and $\N$ is given by Eqs.~(\ref{eq:boltzsol+}),
(\ref{eq:boltzsol0}) and (\ref{eq:liousol-}).

\subsubsection{Perfectly stable vortons with loops production}

As for the Liouville solution, our results also apply to the
particular limit $\gammav \rightarrow 0$. The number of vortons can be
defined as in Eq.~(\ref{eq:calVdef}) and is obtained by integrating
Eq.~(\ref{eq:boltzsol0}) over all loop sizes intercepting the
``0'' domain at time $t$. The first term is the same as in
Eq.~(\ref{eq:vortdensini}) while there is an additional contribution
coming from the loop production function. Defining
\begin{equation}
\tpv(\N) \equiv \dfrac{\ellp(\N) - \ellv(\N)}{\gammad} =
\dfrac{\N}{\gammad} \left(\lambda \N - \dfrac{1}{\sqrt{U}} \right),
\end{equation}
the time required for a loop of size $\ellp(\N)$ to reach $\ellv(\N)$
by gravitational radiation, and the new variable
\begin{equation}
\varsigma \equiv  \dfrac{\ell - \ellv}{\gammav} + t\,,
\end{equation}
one gets
\begin{equation}
\label{eq:vortdens}
\calV(t,\N) = \calVini(t,\N) + \dfrac{2 \ellp}{a^3(t) \N}
\int_{\tini}^t a^3(\varsigma - \tpv) \calP(\ellp,\varsigma - \tpv) \step{\varsigma -
  \tpv} \ud \varsigma\,.
\end{equation}
This expression does no longer depend on $\gammav$ which makes it
valid even for $\gammav \rightarrow 0$. It shows that, in addition to
the vortons coming from the initial loop distribution, loop production
incessantly feeds the vortons reservoir with a time delay given by
$\tpv(\N)$.

The loop distribution when $\gammav \rightarrow 0$, in presence of
a loop production function, finally reads for all $\ell$:
\begin{equation}
\label{eq:vortboltz}
\begin{aligned}
\lim_{\gammav \rightarrow 0} a^3(t) \dfrac{\partial^2 n}{\partial \ell
  \partial \N} (t,\ell,\N) & = \step{\ellv- \ell} \aini^3
\calNini(\ell,\N) \\ & + \step{\ell - \ellv} \aini^3
\calNini\negleft[\ell + \gammad \left( t-\tini \right),\N \right] \\ &
+ \step{\ell - \ellv} \dfrac{2 \ellp}{\gammad \N} a^3\negleft(t -
\dfrac{\ellp - \ell}{\gammad} \right) \calP\negleft(\ellp,t -
\dfrac{\ellp-\ell}{\gammad}\right) \\ & \times \Big\{ \step{\ellp -
  \ell} - \stepb{\ellp - \ell - \gammad(t - \tini)} \Big\} \\ & +
\dirac{\ell -\ellv} \aini^3 \int_{0}^{\gammad(t - \tini)}
\calNini(\ellv + \sigma,\N) \ud \sigma \\ & + \dirac{\ell -\ellv}
\dfrac{2 \ellp}{\N} \int_{\tini}^t a^3(\varsigma - \tpv)
\calP(\ellp,\varsigma - \tpv) \step{\varsigma - \tpv} \ud \varsigma\,.
\end{aligned}
\end{equation}
Note that the initial loop distribution is already responsible
for the so-called vorton excess problem, and we are adding a possibly
larger term to it: as expected, looking in details at the loop distribution
evolution constrains the models even more. We now turn to this
question in the relevant cosmological framework.

\section{Application to the vorton excess problem}

\label{sec:vabund}

In Ref.~\cite{Brandenberger:1996zp}, the relic abundance of vortons was
estimated by assuming that loops are first formed during the string
forming phase transition whereas they develop a current at a later
time. Under some reasonable assumptions,
Ref.~\cite{Brandenberger:1996zp} approximates the number density of
vortons by $n_\uv \propto n_\uf (a_\uf/a)^3$ with $n_\uf
\propto 1/L_\uf^3$ denoting the vorton abundance at formation if their
distribution peaks around a particular length scale $L_\uf$. As we
show below, our Boltzmann treatment allows to derive the complete
current-carrying loop distribution without making assumption on their initial
distribution.

Under the same assumptions as in Ref.~\cite{Brandenberger:1996zp}, we
assume that loops are formed at an energy scale $\Tx \simeq \sqrt{U}$
whereas the current-carrier condensation occurs at the lower
temperature $\Tcur \simeq \lambda^{-1}$. From the results of
Sec.~\ref{sec:evol}, we can readily write down the loop distribution at
the time the strings become superconducting, i.e. at $t=\tcur$:
\begin{equation}
\label{eq:ini2cur}
\barcalN_\ucur(\ell) = \dfrac{\aini^3}{a^3(\tcur)}
\barcalNini[\ell + \gammad(\tcur-\tini)],
\end{equation}
where
\begin{equation}
\barcalN_\alpha \equiv \dfrac{\partial n}{\partial \ell}(\ell,t_\alpha),
\end{equation}
is the loop density distribution at $t=t_\alpha$. In
Eq.~(\ref{eq:ini2cur}), $\gammad$ describes the loop gravitational
decay, complemented or not by any other evaporation effects the loops
may experience in the friction dominated regime. For simplicity, we
will be keeping an unique $\gammad$ in the following. At $t=\tcur$,
current-carrier condensation occurs and those loops are endowed with
the conserved number $N$ induced by thermal fluctuations of wavelength
$\lambda = 1/\Tcur$:
\begin{equation}
\calN_\ucur(\ell,\N) = \barcalN_\ucur(\ell) \,
\dirac{\N-\sqrt{\dfrac{\ell}{\lambda}}}.
\end{equation}
For $\ell > \lambda$, the loops can potentially become stable
vortons. One should nevertheless require that $\ellv(N)=N/\sqrt{U} >
\lambda$, i.e.
\begin{equation}
\N > \Nstar \equiv \lambda \sqrt{U}\,.
\label{eq:Nstar}
\end{equation}
In the opposite situation, $\N<\Nstar$, one may expect the current to
disappear by some quantum processes and these loops are referred to as
``doomed''. This effect can be phenomenologically included within our
interpretation by assigning a vanishing effective vorton length for
those loops, i.e. by imposing $\ellv(\N<\Nstar) = 0$. For the large
enough loops, following the hypothesis of
Ref.~\cite{Brandenberger:1996zp}, there is no loop production and the
vortons having $\ellv>\lambda$ are assumed to be perfectly stable
$\gammav \rightarrow 0$. Under these assumptions,
Eq.~(\ref{eq:vortboltz}) becomes
\begin{equation}
\begin{aligned}
&a^3 \dfrac{\partial^2 n}{\partial \ell \partial
    \N}(t,\ell,\N) = \aini^3\stepb{\ell - \ellv(\N)}
  \barcalNini[\ell+\gammad(t-\tini)] \, \diracb{\N - \sqrt{\dfrac{\ell
        + \gammad(t-\tcur)}{\lambda}}} \\ & +  
  \aini^3 \diracb{\ell - \ellv(\N)} \int_0^{\gammad(t-\tcur)}\negthickspace
  \barcalNini\left[ \ellv(\N) + \sigma + \gammad(\tcur - \tini)\right]
  \, \diracb{\N - \sqrt{\dfrac{\ellv(\N) + \sigma}{\lambda}}} \ud
  \sigma .
\end{aligned}
\label{eq:vexcessall}
\end{equation}
Integrating this expression over $\N$ gives the number density of
proto-vortons plus doomed loops (first line) and vortons (second line)
of size $\ell$ at time $t$:
\begin{equation}
a^3 \dfrac{\partial n}{\partial \ell}(t,\ell) = \int_0^\infty a^3 \dfrac{\partial^2 n}{\partial \ell \partial
    \N}(t,\ell,\N) \,\ud \N.
\end{equation}
Let us start by the number density of vortons. The first Dirac
function in the second line of Eq.~(\ref{eq:vexcessall}) allows to
explicitly perform the integration over $\N$. One has
\begin{equation}
\diracb{\ell - \ellv(\N)} = \sqrt{U} \, \dirac{\N - \ell \sqrt{U}},
\end{equation}
provided $\N>\Nstar$. In the opposite situation, $\ellv(\N)=0$ 
and the integral vanishes. All in all, one can use the above
expression provided $\ell \sqrt{U} > \Nstar$, i.e. for
$\ell > \lambda$ as expected. The vorton term now reads
\begin{equation}
\begin{aligned}
\aini^3 & \step{\ell-\lambda} \sqrt{U} \int_0^{\gammad(t-\tcur)}
\barcalNini\left[\ell + \sigma + \gammad(t-\tcur)\right] \dirac{\ell
  \sqrt{U} - \sqrt{\dfrac{\ell + \sigma}{\lambda}}} \ud \sigma \\
& = 2
\aini^3 \Nstar^2 \dfrac{\ell}{\lambda} \, \step{\ell -\lambda}
\stepb{\ellT(t) - \ell} \, \barcalNini\negleft[ \Nstar^2
  \dfrac{\ell^2}{\lambda} + \gammad (t -\tcur) \right],
\end{aligned}
\label{eq:intsig}
\end{equation}
where $\ellT(t)$ is an increasing function of time defined by
\begin{equation}
\ellT(t) \equiv \dfrac{\lambda}{2 \Nstar^2} \left[1 + \sqrt{1 + 4
    \Nstar^2 \dfrac{\gammad(t-\tcur)}{\lambda}} \right].
\label{eq:ellT}
\end{equation}
As can be checked in Eq.~(\ref{eq:intsig}), the remaining Dirac
function yields a non-vanishing integral provided $\sigma_0 =
\ell(\lambda U \ell-1)$ belongs to the interval
$[0,\gammad(t-\tcur)]$. After some algebra, this ends up being
equivalent to the condition $\ell < \ellT(t)$. Physically, this
condition simply arises from the time required for a loop of size
$\ell$ to shrink down to the vorton length $\ellv$. The quantity
$\ellT(t)$ is also the upper bound of the vorton length spectrum at
any time, their size ranging from $\lambda$ to $\ellT(t)$.

Concerning the first line of Eq.~(\ref{eq:vexcessall}), accounting for
proto-vortons and doomed loops, the integral over $\N$ can again be
performed explicitly owing to the Dirac function. However, one has to
distinguish two cases according to the value of $\N_0=[\ell/ \lambda +
  \gammad(t-\tcur)/\lambda]^{1/2}$, the zero of the Dirac function
argument. Either $\N_0 < \Nstar$ and $\ellv(\N_0) = 0$, or $\N_0 \ge
\Nstar$ and $\ellv(N_0)= N_0/\sqrt{U}$. Again, these conditions can be
recast in terms of lengths by defining a new length scale
\begin{equation}
\ellstar(t) \equiv \lambda \left[ \Nstar^2 - \gammad(t-\tcur)\right],
\label{eq:ellstar}
\end{equation}
which is a decreasing function of time. At any time, all loops having
$\ell < \ellstar(t)$ are doomed, they are the ones associated with $\N
< \Nstar$ and will disappear by decay. For those, we have set above
$\ellv=0$ and the Heaviside function in the first line of
Eq.~(\ref{eq:vexcessall}) equals unity. All the others loops,
having $\ell > \ellstar(t)$ are proto-vortons, i.e. in the shrinking
stage before becoming vortons. This can be explicitly seen by
rewriting the Heaviside function as
\begin{equation}
\stepb{\ell - \ellv(\N_0)} = \stepb{\ellstar(t) - \ell} + \stepb{\ell -
  \ellstar(t)} \stepb{\ell - \ellT(t)}.
\end{equation}
Combining all terms together, the loop distribution function finally
reads
\begin{equation}
\begin{aligned}
a^3(t) \dfrac{\partial n}{\partial \ell}(t,\ell) & = \aini^3 \left\{
\stepb{\ellstar(t) - \ell} + \stepb{\ell - \ellstar(t)} \stepb{\ell -
  \ellT(t)} \right\} \barcalNini\negleft[\ell + \gammad
  (t-\tini) \right] \\ & + 2 \aini^3 \, \Nstar^2 \dfrac{\ell}{\lambda} \,
\stepb{\ellT(t) - \ell} \step{\ell - \lambda}
\barcalNini\negleft[\Nstar^2 \dfrac{\ell^2}{\lambda} +
  \gammad (\tcur - \tini) \right],
\end{aligned}
\label{eq:vabund}
\end{equation}
where $\Nstar$ is given by Eq.~(\ref{eq:Nstar}), $\ellT(t)$ and
$\ellstar(t)$ being defined in Eqs.~(\ref{eq:ellT}) and
(\ref{eq:ellstar}), respectively. As we have just discussed, the
second line of this expression accounts for all vortons present at
time $t$ while the first describe both proto-vortons and doomed
loops. Let us stress that the doomed loops, i.e. the domain $\ell <
\ellstar(t)$ exists only during a transient period after which they
completely disappear. This happens at the time $t=\tstar$ solution of
$\ellstar(\tstar)=0$, i.e. for
\begin{equation}
\tstar = \tcur + \dfrac{\Nstar^2}{\gammad}\,.
\label{eq:tstar}
\end{equation}
The loop distribution of Eq.~(\ref{eq:vabund}) generalizes the
approach of Ref.~\cite{Brandenberger:1996zp} for any initial loop
distributions while tracking at all times both the proto-vorton and
vorton populations. Taking an initial loop distribution peaked at a
particular length gives back the results of
Refs.~\cite{Brandenberger:1996zp, Carter:1999an} in the asymptotic
limit $t \gg \tstar$ and if we neglect the proto-vortons. As a result,
we do not expect significant new constraints to be derived from
Eq.~(\ref{eq:vabund}), i.e. in the case of a vanishing loop production
function. However, as can be seen from Eq.~(\ref{eq:vortboltz}), for
$\calP(\ell,t) \ne 0$, as it is found in the numerical simulations,
the vorton population should be significantly enhanced. We let
however for a future work the derivation of the associated cosmological
constraints.

\section{Conclusion}

We have presented a new way of treating the vorton excess problem,
taking into account in principle any kind of interaction between loops
and long strings, and in particular the appearance of an ensemble of
vortons or quasi-vortons \cite{Brandenberger:1996zp}, i.e. loop
configurations whose internal structure induces a different, possibly
vanishing, decay rate. These vortons could plague GUT models having a
cosmic string network building as a consequence of primordial symmetry
breaking, i.e., basically all cosmologically-compatible GUT
\cite{Davis:1995kk,Jeannerot:2003qv}.  Indeed, in realistic models,
the strings that form are expected to couple to various fields
\cite{Witten:1984eb}, some of which leading to current-carrying
vortices \cite{Peter:1992dw} in such a way that the strings are no
longer Nambu-Goto like as usually assumed
\cite{Carter:1994hn,Martin:1994jp,Martin:2000ca,CorderoCid:2002ts}. This
leads to the formation of very-slowly decaying, or even stable, matter
scaling objects dubbed vortons, that can change the subsequent
evolution in a drastic way. Up to now, only rough evaluation of their
contribution has been proposed using their initial distribution
function.

In this work we have gone one step further into understanding
in more details the vorton distribution and the relevant cosmological
constraints. We have presented a new Boltzmann approach governing the
evolution of the number density $n(t,\ell,\N)$ of current carrying
cosmic string loops. Under some assumptions, we have been able to find
an explicit solution starting from any initial distributions and for
any given loop production functions: the most general results are
given by Eqs.~(\ref{eq:liousol-}), (\ref{eq:boltzsol+}) and
(\ref{eq:boltzsol0}).

We have also shown how this method was
extending previous results on the cosmological evolution of vortons in
the absence of loop production~\cite{Brandenberger:1996zp,
  Carter:1999an}, while remaining perfectly compatible in the
appropriate limits. Our results could be readily applied to the loop
production function found in Ref.~\cite{Ringeval:2005kr,
  Polchinski:2006ee, Dubath:2007mf} to revisit the vorton constraints
in presence of loop production, as should be done in a
forthcoming work. Another possible extension could be the
determination of a complete analytic loop production function but from
a system of Boltzmann equations that would explicitly couple long
strings and loops with collision and fragmentation terms. Indeed, much
still deserves to be done in this area: even though cosmic strings have
long ago been ruled out as the main source of density perturbations,
it does not mean that they ought not be used as a very powerful tool
to provide constraints on theories otherwise unreachable. 

\begin{acknowledgments}
It is a pleasure to thank D.~Steer and M.~Sakellariadou for
enlightening discussions. C.R. is partially supported by the
ESA Belgian Federal PRODEX Grant No.~4000103071 and the
Wallonia-Brussels Federation grant ARC No.~11/15-040.
  
\end{acknowledgments}

\bibliography{strings}

\end{document}